\documentclass[%
 reprint,
 amsmath,amssymb,
 aps,dvipdfmx
]{revtex4-1}
\usepackage{graphicx}
\usepackage{dcolumn}
\usepackage{bm}
\usepackage{color}

\newcommand{\red}[1]{\textcolor{black}{#1}}
\renewcommand{\a}{\alpha}
\renewcommand{\b}{\beta}

\newcommand{\bea}{\begin{eqnarray}}
\newcommand{\eea}{\end{eqnarray}}
\newcommand{\f}[2]{\frac{#1}{#2}}
\newcommand{\eq}{&=&}
\newcommand{\nn}{\nonumber \\ }
\newcommand{\ve}{\varepsilon}
\renewcommand{\d}{\delta}
\newcommand{\area}{\int_{-\infty}^\infty }

\newcommand{\p}{\partial}
\newcommand{\pp}[2]{\f{\p #1}{\p #2}}

\newcommand{\sref}[1]{Eq. (\ref{#1})}

\newcommand{\rx}{\tau}

\newcommand{\citeauthorname}[2]{{#1} {#2}}
\newcommand{\citebook}[4]{{#1} {\it #2} ({#3}, {#4}).}
\newcommand{\citepaper}[4]{{#1} {#3} ({#4}).}
\begin{document}

\preprint{APS/123-QED}

\title{Minimal Investment Risk of Portfolio Optimization Problem\\ with 
Budget and Investment Concentration Constraints}

\author{Takashi Shinzato}
\email{takashi.shinzato@r.hit-u.ac.jp}
 \affiliation{
Mori Arinori Center for Higher Education and Global Mobility,
Hitotsubashi University, 
Tokyo, 1868601, Japan.}

\date{\today}

\begin{abstract}
In the present paper, 
the minimal investment risk for a portfolio optimization problem 
with imposed budget and investment concentration constraints 
is considered using replica analysis.
Since the minimal investment risk is influenced by the investment concentration constraint (as well as the budget constraint), 
it is intuitive that 
the minimal investment risk for the problem with an investment concentration constraint be larger than that without the constraint 
(that is, with only the budget constraint). Moreover, 
a numerical experiment shows the effectiveness of our proposed analysis.
\begin{description}
\item[PACS number(s)]
{89.65.Gh}, {89.90.+n}, {02.50.-r}
\end{description}
\end{abstract}
\pacs{89.65.Gh}
\pacs{89.90.+n}
\pacs{02.50.-r}
\maketitle


\section{Introduction}
Portfolio optimization problems constitute one of the most important themes
in the research field of mathematical finance and 
first appeared in  
the theory of diversification investment introduced by Markowitz in 
1952. Analytical procedures for solving 
portfolio optimization problems 
to accurately implement asset management 
so as to disperse the risk by diversifying investment into several 
assets have become well known  
\cite{Markowitz1952,Markowitz1959,Bodie,Luenberger}.
Over the next few decades, 
several issues related to portfolio optimization problems have been 
{addressed}, and recently 
several models and the behavior of the minimal investment risk in 
portfolio optimization problems have 
been thoroughly examined using {analytical approaches 
which have }  
been developed and improved {through multidisciplinary collaboration} \cite{Ciliberti,Kondor,Pafka,Shinzato-SA2015,Shinzato-BP2015}.
Taking advantage of such {earlier} works, 
we can find the behavior of the optimal solution 
for an optimization problem with respect to stochastic phenomena; for 
instance, we can assess the learning performance of perceptron learning and 
can evaluate the performance of the decoding algorithm of low-density parity-check code and 
code division multiple access  
using an analytical approach developed in 
statistical mechanical informatics 
\cite{Shinzato2008,Kabashima2004,Tanaka2002}. 
Obviously, 
since 
portfolio optimization problems are 
formulated as stochastic problems, 
that is, 
modeled within the framework of probabilistic inference, 
in order to investigate the behavior of the optimal portfolio, 
it is useful to adopt an analytical method whose effectiveness has been verified 
in several research fields. 
For instance, 
Ciliberti et al. analyzed 
the minimal investment risk under 
an absolute deviation model and an expected shortfall model using 
replica analysis in the absolute zero temperature limit \cite{Ciliberti}. 
Kondor et al. examined
quantitatively the noise sensitivity of the optimal portfolio for several risk functions 
\cite{Kondor}. Pafka et al. 
discussed the relation between predicted risk, realized risk, and true risk in detail
via a scenario ratio (between the number of scenarios and the number of 
assets) \cite{Pafka}. \red{
Shinzato derived the statistics minimal investment risk and investment 
concentration and showed that the minimal investment risk is attained 
and the investment concentration constraint is satisfied (e.g., by a 
portfolio) and that these two statistics have the self-averaging 
property which is frequently used in statistical mechanical informatics 
analysis \cite{Shinzato-SA2015}.} In addition, 
Furthermore, Shinzato et al. developed
a faster algorithm for solving portfolio optimization problems 
using a belief propagation method \cite{Shinzato-BP2015}.

However, 
this previous work mainly investigated 
the minimal investment risk of the portfolio optimization problem 
with only one constraint, for instance, a budget constraint. 
Therefore, we need to 
analyze in detail the minimal investment risk of a portfolio optimization problem 
with several constraints, where these constraints correspond to several policies of 
the stakeholder, 
so as to handle more practical situations.
As the first step of 
analyzing portfolio optimization problems with several constraints, 
noting the risk minimization problem, 
which makes it possible to 
characterize the investment strategy of a risk-averse investor, 
our purpose in this {study is to develop} 
a novel approach for solving 
a portfolio optimization 
problem with two representative constraints, a budget constraint and a constraint on 
{investment concentration}, as well as to assess theoretically 
the minimal investment risk and {investment concentration} 
using replica analysis. In {these previous works}, 
the minimal investment risk and the {investment concentration} of the portfolio optimization problem with only a
budget constraint are discussed in detail. However, 
we can also consider an {investment concentration} constraint, since we assess
quantitatively how the minimal investment risk is influenced by the 
{investment concentration}, which can be expected to lead to new findings regarding risk-averse investors.

This paper is organized as follows: In the next section, 
we {prepare} one of the most {analyzed} models with respect to portfolio optimization 
problems, the mean-variance model, \red{by formulating it} with two
constraints, a budget constraint and an {investment concentration} constraint.
In section \ref{sec3}, replica analysis is applied to the 
portfolio optimization problem with these two constraints, 
followed by an analytical procedure used in previous work.
In section \ref{sec4}, in order to 
verify the effectiveness of our proposed approach, 
we compare the results derived using the proposed method with 
those from numerical simulation and those obtained using the 
standard approach in operations research. 
The final section is devoted to summarizing the paper and discussing possible future work.

\section{Mean-Variance Model with Two Constraints}
This paper considers optimally diversified investment in $N$ assets in an investment market 
with no restrictions on short selling. Herein, $w_i$ is a the amount of asset $i(=1,\cdots,N)$ in the portfolio and 
the full portfolio of $N$ assets is denoted $\vec{w}=\left\{w_1,\cdots,w_N\right\}^{\rm T}\in{\bf R}^N$, where ${\rm T}$
indicates the transpose of a vector or matrix. 
$x_{i\mu}$ is the return rate of asset $i$ under scenario 
$\mu(=1,\cdots,p)$. 
For simplicity of our discussion, {similar to in the previous work}, 
it is assumed that each return rate $x_{i\mu}$ is independently and identically {normally}  distributed 
with mean $0$ and variance $1$ \cite{Shinzato-SA2015}.
Under this assumption, 
given the $p$ return rate vectors 
$\vec{x}_1,\cdots,\vec{x}_p\in{\bf 
R}^N,\vec{x}_\mu=\left\{x_{1\mu},\cdots,x_{N\mu}\right\}^{\rm T}\in{\bf 
R}^N$, written in matrix notation as the return rate matrix 
$X=\left\{\f{x_{i\mu}}{\sqrt{N}}\right\}\in{\bf R}^{N\times p}$, 
in the mean-variance model, 
the investment risk ${\cal H}(\vec{w}|X)$ of portfolio $\vec{w}$ is defined 
as follows:
\bea
\label{eq1}
{\cal H}(\vec{w}|X)\f{1}{2}
\sum_{\mu=1}^p
\left(\f{1}{\sqrt{N}}\sum_{i=1}^Nw_ix_{i\mu}\right)^2.
\eea
Note that 
the necessary and sufficient condition 
for the optimal portfolio for portfolio optimization problem to be uniquely determined 
given in {\cite{Shinzato-SA2015}}
 is that $J=XX^{\rm T}\in{\bf R}^{N\times N}$ be a non-singular 
 matrix, that is, the rank of matrix $J$ be $N$ or simply $p>N$.
However, since $J$ does not always need to be a regular 
matrix 
to guarantee a unique optimal portfolio for a portfolio optimization problem with several 
constraints, as 
in the present work, we do not adopt the regular matrix assumption here.
Moreover, coefficient $1/\sqrt{N}$
 guarantees that the statistics defined below are significant, 
since the expectation of the return rate is assumed to be $0$ and the 
expected return  of portfolio $\vec{w}$ is regarded as $0$, 
we {omit} 
terms of
expected return. Furthermore, as one of the constraints, the budget 
constraint is as follows:
\bea
\label{eq2}
\sum_{i=1}^Nw_i\eq N.
\eea
Note that this constraint differs from the
budget constraint in the standard context of operations research, 
since 
the investment ratios between asset $i$ and asset $j$ in optimal portfolios 
derived from both budget constraints are consistent with each other 
and we can rescale the budget constraint satisfied with  those statistics discussed hereafter for characterizing the investment
style of the optimal 
portfolio, {we employ} \sref{eq2}.
For further details about the budget constraint, please refer to {\cite{Shinzato-SA2015}.}

If the portfolio which minimizes the investment risk function in 
\sref{eq1} under the budget constraint in \sref{eq2} is represented as 
$\vec{w}^*=\left\{w_1^*,\cdots,w_N^*\right\}^{\rm T}={\rm 
arg}{\min}_{\vec{w}}{\cal H}(\vec{w}|X)$, 
from a finding in {\cite{Shinzato-SA2015}},  
it is guaranteed that 
the minimal investment risk per asset $\ve$ is attained, 
its  {investment concentration} $q_w$ 
satisfies the investment risk constraint, and these two statistics have the self-averaging property  as $N$ and $p$ approach 
infinity while keeping 
$p/N$ bounded {\cite{Shinzato-SA2015,Shinzato2008,Kabashima2004,Tanaka2002}}. Specifically, since ${\cal H}(\vec{w}^*|X)=E_X[{\cal H}(\vec{w}^*|X)]$ and 
$\sum_{i=1}^N(w_i^*)^2=\sum_{i=1}^NE_X[(w_i^*)^2]$, where 
$E_X[f(X)]$ means the expectation of $f(X)$ with respect to return rate matrix $X$,
the minimal investment risk per asset and 
the {investment concentration} are given analytically as follows:
\bea
\label{eq3}
\ve\eq
\f{1}{N}{\cal H}(\vec{w}^*|X)\nn
\eq\left\{
\begin{array}{ll}
\f{\a-1}{2}&\a>1\\
0&\text{otherwise}
\end{array}
\right.,\\
\label{eq4}
q_w\eq
\f{1}{N}\sum_{i=1}^N(w_i^*)^2\nn
\eq\left\{
\begin{array}{ll}
\f{\a}{\a-1}&\a>1\\
\infty&\text{otherwise}
\end{array}
\right.,
\eea
in which $\a=p/N\sim O(1)$ is the scenario ratio. 
 
Therefore, as is intuitive,
the optimal portfolio 
$\vec{w}^*$ that minimizes investment risk 
${\cal H}(\vec{w}|X)$ {in} \sref{eq1}, which is defined for a given return rate 
matrix $X$, {depends on return rate matrix} $X$. However, 
\red{since the minimal investment risk is attained, the investment concentration satisfies its constraint, and both have the self-averaging property},  
one can accurately assess their values using replica analysis 
\red{more easily than using numerical simulations}.  In contrast, 
in operations research, 
based on the analytical approach derived from the principle of expected 
utility maximization, 
one solves for the portfolio which minimizes the expected investment risk $E_X[{\cal 
H}(\vec{w}|X)]$, $\vec{w}^{* \rm OR}=\left\{w_1^{*\rm OR},\cdots,w_N^{*\rm 
OR}\right\}^{\rm T}={\rm arg}\mathop{\min}_{\vec{w}}E_X[{\cal 
H}(\vec{w}|X)]$, and then the minimal expected investment risk per asset 
$\ve^{\rm OR}$ and its {investment concentration} $q_w^{\rm OR}$ can 
be \red{estimated briefly} as follows \cite{Shinzato-SA2015}:
\bea
\label{eq5}
\ve^{\rm OR}
\eq\f{1}{N}E_X[{\cal H}(\vec{w}^{*\rm OR}|X)]\nn
\eq
\f{\a}{2},\qquad \a>0,\\
\label{eq6}
q_w^{\rm OR}\eq
\f{1}{N}\sum_{i=1}^N(w_i^{*\rm OR})^2\nn
\eq1,\qquad \a>0.
\eea
From \sref{eq3} and 
\sref{eq5},
$\ve<\ve^{\rm OR}$ is obtained, 
{under appropriately realistic conditions}, the portfolio which minimizes the expected investment risk $\vec{w}^{*\rm 
OR}$ derived using the standard operations research approach 
cannot minimize ${\cal H}(\vec{w}|X)$ characterized by {an arbitrary} return 
rate matrix $X$. {Therefore, it is necessary to} estimate properly 
the optimal investment strategy {that can minimize the }investment 
risk  with respect to {a return rate matrix}. Finally, 
from the previous argument,  the analytical approach widely used in operations research is 
regarded as an annealed disordered system approach, which 
has already been shown in previous interdisciplinary {works} to be 
impractical due to the impossibility of analyzing the quenched 
disordered system and of evaluating the expectation of the {optimal} 
objective 
function {\cite{Shinzato-SA2015,Shinzato-BP2015,Shinzato2008,Kabashima2004,Tanaka2002}}. 
That is, the approach of annealed disordered systems can provide little valuable knowledge regarding investing to investors.

In contrast, it has already been shown in the interdisciplinary 
research {\cite{Shinzato-SA2015,Shinzato-BP2015,Shinzato2008,Kabashima2004,Tanaka2002}} that the quenched disordered system approach  can easily be used 
to analyze {a quenched disordered system and evaluate
the expectation of an} {optimal objective function and} 
we can evaluate 
{the inherent investment risk of an investment system} 
using the quenched disordered system approach and 
obtain a variety of valuable knowledge and several optimal investment 
strategies for investors. 
To achieve the above, as a first step toward precisely analyzing 
a given investment system which {satisfies the definition of} a 
quenched disordered system, we must investigate {the potential risk 
of an 
investment system}.

In a previous study \cite{Shinzato-SA2015}, 
the minimal investment risk and its {investment concentration} of a portfolio optimization problem with only a budget
constraint were analyzed. Therefore, in the present paper, 
we consider the two-constraint case; namely, the following novel constraint is added:
\bea
\label{eq7}
\f{1}{N}\sum_{i=1}^Nw_i^2\eq \rx,
\eea
where 
$\rx$ is a scalar constant. This constraint implies that the {investment concentration} defined in 
\sref{eq4} is {held} constant. 
Thus, the portfolio optimization problem discussed in the previous work 
is  extended 
to the optimization problem of finding $\vec{w}$ that minimizes the investment risk in \sref{eq1} under the budget constraint in \sref{eq2} 
and the {investment concentration} constraint in \sref{eq7}.

Let us describe the {investment concentration} before discussing 
this portfolio optimization problem. It is easily understood through 
a comparison between two investment strategies: (concentrated investment 
strategy: CIS) the investor invests in asset $1$ only, that is, 
$\vec{w}^{\rm CIS}=\left\{N,0,\cdots,0\right\}^{\rm T}\in{\bf R}^N$; and
(equipartition investment strategy: EIS) the investor invests 
equally in $N$ assets, that is, $\vec{w}^{\rm 
EIS}=\left\{1,\cdots,1\right\}^{\rm T}\in{\bf R}^N$. 
Then
$q_w^{\rm CIS}(=N)>q_w^{\rm EIS}(=1)$. In general, 
the larger the {investment concentration} is, the fewer the assets the investor 
{tends to invest in. That is, }
under the {investment concentration} and budget constraints, 
we should determine the optimal portfolio among the set of portfolios whose 
{investment concentrations} are $\rx$, 
{which is equivalent to modeling the case of preventing overconcentration in investing}. 
 
\sref{eq2} and \sref{eq7} imply
\bea
\label{eq8}
\rx-1\eq
\f{1}{N}\sum_{i=1}^Nw_i^2-
\left(\f{1}{N}\sum_{i=1}^Nw_i\right)^2\nn
\eq\f{1}{N}\sum_{i=1}^N
\left(w_i-\f{1}{N}\sum_{i=1}^Nw_i\right)^2.
\eea
Thus, the constant $\rx$ in \sref{eq7} is at least 
$1$. Moreover, we apply the budget 
constraint in \sref{eq2} rather than $\sum_{i=1}^Nw_i=1$ as widely used in 
operations research, because the latter constraint cannot be satisfied 
simultaneous with the relation  \sref{eq8}, making {the constraints}  hard to 
interpret in the context of statistics.  
As in previous studies \cite{Shinzato-SA2015,Ciliberti,Pafka}
we normalize the return rate to mean $0$, as described previously. 
Because of this, the sum of the expected 
returns of the assets of the portfolio is also 0 and cannot be used as a 
constraint. Therefore, 
we instead constrain the {investment concentration}.

Our aim in this work is to compare the minimal investment risk for the 
budget constraint in \sref{eq2} with 
that with 
both the budget constraint in \sref{eq2} and the {investment concentration} constraint in \sref{eq7}.
A further aim is to {compare} the minimal investment risk under the influence of {added constraints.}

\section{Replica Analysis\label{sec3}}
In a similar way to that used in the previous work {\cite{Shinzato-SA2015}}, we also 
employ 
replica analysis as developed in 
statistical mechanical informatics 
so as to analyze the 
minimal investment risk in the portfolio optimization problem 
with constraints \sref{eq2} and \sref{eq7}. If 
the investment risk in the market ${\cal H}(\vec{w}|X)$
is regarded as the Hamiltonian of the canonical ensemble,  then
the partition function of inverse temperature $\b$ is defined as follows:
\bea
Z(X)=\area d\vec{w}\d(\vec{w}^{\rm T}\vec{e}-N)
\d(\vec{w}^{\rm T}\vec{w}-N\rx)
e^{-\b{\cal H}(\vec{w}|X)},\qquad
\eea
where $\vec{e}=\left\{1,\cdots,1\right\}^{\rm T}\in{\bf R}^N$ and constraints
\sref{eq2} and \sref{eq7} are handled using the delta function.

For a statistical mechanical informatics scenario, 
we need to find the Helmholtz free energy per asset (or free entropy per asset \cite{Mezard}) 
in a quenched disordered system in order to 
calculate the minimal investment risk per asset. For the present study, 
similar to in previous work \cite{Shinzato-SA2015},
since it is difficult to directly implement configuration averaging 
of the logarithm of the partition function, 
we analyze this problem 
using configuration averaging of {a power} of the partition function
and the replica trick \cite{Nishimori}. Here we define 
{the following order parameters:}
\bea
\label{eq10}
q_{wab}\eq\f{1}{N}\sum_{i=1}^Nw_{ia}w_{ib},\qquad(a,b=1,\cdots,n).
\eea
Moreover, conjugate order parameters $\tilde{q}_{wab}$ are 
prepared and {we assume the following replica symmetry solution:}
\bea
\label{eq11}
q_{wab}
\eq\left\{
\begin{array}{ll}
\chi_w+q_w&a=b\\
q_w&a\ne b
\end{array}
\right.,\\
\tilde{q}_{wab}
\eq\left\{
\begin{array}{ll}
\tilde{\chi}_w-\tilde{q}_w&a=b\\
-\tilde{q}_w&a\ne b
\end{array}
\right.,\\
k_a\eq k,\\
\theta_a\eq\theta,
\eea
where $k_a$ and $\theta_a$  are the conjugate variables of the constraint conditions in \sref{eq2}
and 
\sref{eq7}, respectively. Next, we define
\bea
\phi\eq\lim_{N\to\infty}\f{1}{N}
E_X
\left[\log Z(X)\right]\nn
\eq
\mathop{\rm Extr}_{k,\theta,\chi_w,q_w,\tilde{\chi}_w,\tilde{q}_w}
\left\{
-k-\f{\rx\theta}{2}+\f{(\chi_w+q_w)(\tilde{\chi}_w-\tilde{q}_w)}{2}
\right.
\nn
&&+\f{q_w\tilde{q}_w}{2}-\f{1}{2}\log(\tilde{\chi}_w-\theta)+\f{\tilde{q}_w+k^2}{2(\tilde{\chi}_w-\theta)}\nn
&&
\left.
-\f{\a}{2}\log(1+\b\chi_w)-\f{\a\b q_w}{2(1+\b\chi_w)}
\right\},
\eea
where $\mathop{\rm Extr}_mg(m)$ means 
the extremum of $g(m)$ with respect to $m$. Now, the saddle point 
equation of the parameters  is obtained as follows:
\bea
k\eq\tilde{\chi}_w-\theta,\\
\theta\eq \tilde{\chi}_w-\f{1}{\chi_w},\\
\label{eq18}
\chi_w\eq \rx-q_w,\\
q_w\eq\chi_w^2\tilde{q}_w+1,\\
\tilde{\chi}_w\eq\f{\a\b}{1+\b\chi_w},\\
\tilde{q}_w\eq\f{\tilde{\chi}_w^2}{\a}q_w.
\eea
Solving these in the limit of large $\b$, we have 
\bea
\chi_w\eq
\left\{
\begin{array}{ll}
\f{1}{
\b\left(
\sqrt{\f{\a \rx}{\rx-1}}-1
\right)
}&1-\f{1}{\rx}\le\a\\
\rx-\f{1}{1-\a}&\text{otherwise}
\end{array}
\right.
,\\
q_w\eq 
\left\{\begin{array}{ll}
\rx-\f{1}{
\b\left(
\sqrt{\f{\a \rx}{\rx-1}}-1
\right)
}&1-\f{1}{\rx}\le\a\\
\f{1}{1-\a}&\text{otherwise}
\end{array}
\right.
.
\eea
For both cases of $\a$, from \sref{eq4}, \sref{eq10}, and 
\sref{eq11} with $a=b$, the {investment concentration} of the optimal 
portfolio is 
\bea
\label{eq24-1}
\lim_{N\to\infty}\f{1}{N}
\sum_{i=1}^N(w_i^*)^2
\eq \chi_w+q_w\nn
\eq\rx,
\eea
{which also satisfies \sref{eq7}; alternatively, \sref{eq24-1} can be derived 
directly from \sref{eq18}}. Moreover, 
following the statistical mechanical informatics literature, 
if the minimal investment risk per asset $\ve$ is 
determined from 
$\ve=\lim_{\b\to\infty}\left\{-\pp{\phi}{\b}\right\}$, then
\bea
\label{eq24}
\ve\eq
\lim_{\b\to\infty}
\left\{
\f{\a\chi_w}{2(1+\b\chi_w)}+\f{\a q_w}{2(1+\b\chi_w)^2}
\right\}
\nn
\eq\left\{
\begin{array}{ll}
\f{\a \rx+\rx-1-2\sqrt{\a \rx(\rx-1)}}{2}&1-\f{1}{\rx}\le\a\\
0&\text{otherwise}
\end{array}
\right.
.
\label{eq24}
\eea
If $1-\f{1}{\rx}\le\a$, 
since the numerator in \sref{eq24} can be factored as 
$\a \rx+\rx-1-2\sqrt{\a \rx(\rx-1)}=(\sqrt{\a \rx}-\sqrt{\rx-1})^2$,
it is clear that the minimal investment risk per asset is always non-negative. 
Furthermore, if one shifts $\ve$ to 
\bea
\ve\eq\f{\a-1}{2}+\f{(\sqrt{\a(\rx-1)}-\sqrt{\rx})^2}{2},
\eea
then $\ve$ is guaranteed to be greater than \sref{eq3}, 
the minimal investment risk for the portfolio optimization 
problem with \sref{eq7}. 
On the other hand, in the case of $\sqrt{\a(\rx-1)}-\sqrt{\rx}=0$, 
$\rx=\a/(\a-1)$, 
as in \sref{eq7}, 
and the problem can  be regarded as the portfolio optimization problem 
with a budget constraint.

Using the standard operations research approach, 
in a similar way to \sref{eq5} and \sref{eq6}, 
the minimal expected investment risk per asset of the solution 
to the portfolio optimization problem with both constraints, \sref{eq2} and \sref{eq7}, 
$\ve^{\rm OR}$ and its {investment concentration} $q_w^{\rm OR}$ are 
easily  derived as follows:
\bea
\label{eq26}
\ve^{\rm OR}
\eq\f{\a \rx}{2},\\
\label{eq27}
q_w^{\rm OR}\eq \rx.
\eea
If $\a>1-\f{1}{\rx},\rx\ge1$, then $\rx-1-2\sqrt{\a \rx(\rx-1)}<0$ holds;
if $0<\a\le1-\f{1}{\rx},\rx\ge1$, then $\a\rx>0$ also holds, 
that is, the 
minimal investment risk per asset $\ve$ is smaller than the minimal 
expected investment risk per asset $\ve^{\rm OR}$,
\bea
\ve&<&\ve^{\rm OR}.
\eea
Further, from \sref{eq6}, $\rx=1$ is confirmed, 
so that \sref{eq26} and \sref{eq27} {agree with}  the analytical findings, 
\sref{eq5} and \sref{eq6}, 
in the case of a portfolio optimization problem with a budget constraint 
only.

Moreover, with respect to 
the result derived  from our proposed approach, 
instead of the {investment concentration} constraint, \sref{eq7}, 
\bea
\label{eq29}
\rx_0&\le&\f{1}{N}\sum_{i=1}^Nw_i^2,
\eea
{that is, one} determines the portfolio which minimizes the investment risk of \sref{eq1} 
in the set of portfolios whose {investment concentrations} are larger 
than constant $\rx_0$. If $\a>1$, 
then the minimal investment risk per asset $\ve(\rx_0)$ and its {investment concentration} $q_w(\rx_0)$ are as follows:
\bea
\ve(\rx_0)
\eq
\mathop{\min}_{\rx_0\le \rx}
\left[
\f{\a \rx+\rx-1-2\sqrt{\a \rx(\rx-1)}}{2}
\right]
\nn
\eq
\left\{
\begin{array}{ll}
\f{\a-1}{2}&1\le \rx_0<\f{\a}{\a-1}\\
\f{\a \rx_0+\rx_0-1-2\sqrt{\a \rx_0(\rx_0-1)}}{2}&\f{\a}{\a-1}\le \rx_0
\end{array}
\right.,\nn\\
q_w(\rx_0)
\eq
\left\{
\begin{array}{ll}
\f{\a}{\a-1}&1\le \rx_0<\f{\a}{\a-1}\\
\rx_0&\f{\a}{\a-1}\le \rx_0
\end{array}
\right..
\eea
If $0<\a\le1$, we have
\bea
\ve(\rx_0)
\eq0\qquad1\le\rx_0,\\
q_w(\rx_0)&\ge&\f{1}{1-\a}\qquad1\le\rx_0.
\eea
In a similar way, for the case of 
\bea
\label{eq32}
\rx_0&\ge&
\f{1}{N}\sum_{i=1}^Nw_i^2,
\eea
then  if $\a>1$, $\ve(\rx_0)$ and $q_w(\rx_0)$ are
\bea
\ve(\rx_0)
\eq
\mathop{\min}_{\rx_0\ge \rx}
\left[
\f{\a \rx+\rx-1-2\sqrt{\a \rx(\rx-1)}}{2}
\right]
\nn
\eq
\left\{
\begin{array}{ll}
\f{\a \rx_0+\rx_0-1-2\sqrt{\a \rx_0(\rx_0-1)}}{2}
&1\le \rx_0<\f{\a}{\a-1}\\
\f{\a-1}{2}
&\f{\a}{\a-1}\le \rx_0
\end{array}
\right.,\nn\\
q_w(\rx_0)
\eq
\left\{
\begin{array}{ll}
\rx_0&1\le \rx_0<\f{\a}{\a-1}\\
\f{\a}{\a-1}&\f{\a}{\a-1}\le \rx_0
\end{array}
\right.,
\eea
whereas if $0<\a\le1$, they are given by
\bea
\ve(\rx_0)\eq
\left\{
\begin{array}{ll}
\f{\a\rx_0+\rx_0-1-2\sqrt{\a\rx_0(\rx_0-1)}}{2}&1\le\rx_0<\f{1}{1-\a}\\
0&\f{1}{1-\a}\le\rx_0
\end{array}
\right.,\nn\\
q_w(\rx_0)
\eq
\left\{
\begin{array}{ll}
\rx_0&1\le \rx_0<\f{1}{1-\a}\\
\f{1}{1-\a}+c\qquad\ &\f{1}{1-\a}\le \rx_0
\end{array}
\right.,
\eea
for any $c\ge0$. On the other hand, 
if considered in the context of operations research, 
with respect to both optimization problems, that is, those with the {investment concentration} constraint modified to \sref{eq29} or \sref{eq32}, 
the minimal expected investment risk per asset $\ve^{\rm OR}(\rx_0)$ and 
its {investment concentration} $q_w^{\rm OR}(\rx_0)$ are easily
determined as follows:  
\bea
\ve^{\rm OR}(\rx_0)\eq\f{\a \rx_0}{2},\\
q_w^{\rm OR}(\rx_0)\eq \rx_0,
\eea
for which $\ve(\rx_0)<\ve^{\rm OR}(\rx_0)$ holds. 
Namely, we can verify intuitively the findings derived from replica analysis 
{for these two problems}.

\section{Numerical Experiment\label{sec4}}

In order to verify our proposed approach based on 
the assumption of a replica symmetry solution and 
the thermodynamic limit of the number of assets or scenarios, 
we should assess numerically the portfolio which minimizes the investment risk for 
the optimization problem with the constraints \sref{eq2} and \sref{eq7} 
using the Lagrange undetermined multipliers method, 
which can solve for the optimal solution without these assumptions
and compare the analytical results via numerical simulation. 
The Lagrange function is defined as follows:
\bea
L(\vec{w},k,\theta)
=\f{1}{2}\vec{w}^{\rm T}J\vec{w}+k(N-\vec{w}^{\rm T}\vec{e})
-\f{\theta}{2}(\vec{w}^{\rm T}\vec{w}-N\rx),\qquad
\eea
where the $i,j$ component of variance-covariance matrix $J(=XX^{\rm 
T})=\left\{J_{ij}\right\}\in{\bf R}^{N\times N}$ is 
\bea
J_{ij}\eq\f{1}{N}\sum_{\mu=1}^px_{i\mu}x_{j\mu}.
\eea

We optimize Lagrange function 
$L(\vec{w},k,\theta)$ 
by using the following algorithm based on the steepest descent method.
Given a return rate matrix $X=\left\{\f{x_{i\mu}}{\sqrt{N}}\right\}\in{\bf R}^{N\times p}$,
the initial states of portfolio $\vec{w}$ and two conjugate variables 
$k,\theta$ are adequately {initialized, for instance, 
$\vec{w}^0=\vec{e}$ and $k^0=\theta^0=1$}. 
 At iteration step $s$, $\vec{w}^s,k^s,\theta^s$ are updated as follows:
\bea
\vec{w}^{s+1}\eq\vec{w}^s-\eta_w\pp{L(\vec{w}^s,k^s,\theta^s)}{\vec{w}^s},\\
k^{s+1}\eq k^s+\eta_k\pp{L(\vec{w}^s,k^s,\theta^s)}{k^s},\\
\theta^{s+1}\eq \theta^s+\eta_\theta\pp{L(\vec{w}^s,k^s,\theta^s)}{\theta^s},
\eea
where learning steps $\eta_w,\eta_k,\eta_\theta$ are infinitesimal 
positive numbers, set as $\eta_w=\eta_k=\eta_\theta=10^{-1}$ in the 
present experiment. Moreover, when the difference 
$\Delta=\sum_{i=1}^N|w_i^s-w_i^{s+1}|+|k^s-k^{s+1}|+|\theta^s-\theta^{s+
1}|$ is less than $10^{-4}$, then numerical solution $\vec{w}^s$ is 
regarded as an approximation solution of the optimal portfolio 
$\vec{w}^*={\rm arg}\mathop{\min}_{\vec{w}}{\cal H}(\vec{w}|X)$,
and is used to  
estimate the minimal investment risk per asset $\ve$.

\begin{figure}[tb]
\begin{center}
\includegraphics[width=0.9\hsize]{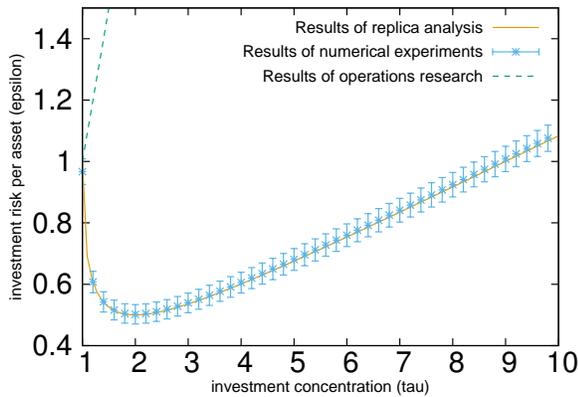}
\caption{
\label{concentrated-investment-level-fixed}
\label{Fig1}
Minimal investment risk per asset $\ve$ {at $\a=p/N=2$} results from replica 
 analysis (orange line), numerical simulation ({sky-blue} asterisks 
 {with error bars}), 
and operations research approach (green dashed line) versus 
 {investment concentration} $\rx$.
Results of replica analysis  are {consistent with} 
the averages 
obtained from a numerical experiment 
 with $100$ samples and $N=500$ assets. 
}
\end{center}

\end{figure}

The number of assets in a numerical simulation $N$ is set as $N=500$ and 
the scenario ratio $\a=p/N$ is set as $\a=2$. Furthermore, 
we set return rate of each asset $x_{i\mu}$ as independently and identically 
distributed following the standard normal distribution $N(0,1)$, 
and 100 return rate matrices $X^1,\cdots,X^{100}$ are prepared  as {sample sets}.
We assess the average of minimal investment risk per asset 
using the optimal portfolio of each sample, $\vec{w}^{*,1},\cdots,\vec{w}^{*,100}$, 
where $\vec{w}^{*,c}={\rm arg}\mathop{\min}_{\vec{w}}{\cal 
H}(\vec{w}|X^c)$ is obtained using the above-described Lagrange undetermined multipliers 
algorithm. 
In Fig. \ref{concentrated-investment-level-fixed}, 
the analytical results derived by replica analysis (\sref{eq24}) 
are shown with the analytical results obtained by the standard operations research approach 
 (\sref{eq26}) and the practical  
results estimated by numerical experiments for purposes of comparison. 
As shown, 
the results derived using replica analysis are consistent 
with those estimated using numerical simulation, whereas the results 
obtained using the standard operations research approach 
{are not. This} 
implies that 
the portfolio derived by the approach widely used in operations 
research based on the principle of expected utility maximization, 
which minimizes the expected investment risk, $\vec{w}^{*\rm OR}$, 
is not always able to minimize {the investment risk}  ${\cal 
H}(\vec{w}|X)$. Thus, the operations research approach may not be able to provide 
the optimal investment strategy desired by investors.

\section{Summary and Future Work}

In the present paper, we have discussed the minimal investment risk per 
asset 
for a portfolio optimization problem 
with two imposed constraints, a budget constraint and an {investment concentration} constraint, using replica analysis, which was developed for
and improved during interdisciplinary research.
Unlike the minimal investment risk per asset and portfolio optimization problem with 
a budget constraint which has been discussed in previous work, 
we assessed quantitatively the deviation  
of the minimal investment risk per asset from the budget 
constraint only case caused by the inclusion of an {investment 
concentration} constraint. Moreover, 
we could estimate the typical behavior of minimal investment risk per asset 
using the Lagrange method of undetermined multipliers,
which is an algorithm for finding the optimal portfolio.
The results obtained using the proposed method, those 
obtained by a standard operations research approach, and numerical results were compared. 
We found that the numerical simulation results were consistent with those of 
replica analysis. In contrast, 
the standard operations research approach failed to {identify}  
accurately the minimal investment risk of the portfolio 
optimization problem, since the obtained optimal portfolio only minimizes the expected investment 
risk, not {the investment risk}, making it clear that 
this approach cannot provide investors information about 
the optimal investment strategy.

For simplicity of discussion, 
we have assumed, in a similar way to in previous work, that the return rate of each asset is normalized as having mean $0$ and variance 
$1$. However, in future work,
we need to improve and develop the model 
in order to be able to treat a more realistic depiction of the investment 
market. For instance, we need to analyze 
the portfolio optimization problem 
in an investment market comprising a risk-free asset and assets 
of varying risk levels. In addition, 
as alternative constraints to a budget constraint or an {investment concentration} constraint, 
we need to consider, for instance, 
an expected return constraint for the case that the return rate is not 
normalized and linear inequality constraints. For such research, since there
has been little investigation based on statistical mechanical informatics of 
the various issues related to portfolio optimization problems, 
 there are {several unresolved issues in 
this researching field}.

\section*{Acknowledgements}
The author appreciates the fruitful comments of I. Arizono, 
H. Nakamura, T. Misumi, H. Yamamoto, and K. Kobayashi. This work is partly supported
by Grant-in-Aid No. 15K20999; the President Project for Young Scientists at Akita Prefectural
University; research project No. 50 of the National Institute of Informatics, Japan; research project No. 5 of
the Japan Institute of Life Insurance; research project
of the Institute of Economic Research Foundation at Kyoto University; research project No. 1414 of the Zengin
Foundation for Studies in Economics and Finance; research project No. 2068 of the Institute of Statistical
Mathematics; research project No. 2 of the Kampo
Foundation; and research project of the Mitsubishi UFJ Trust Scholarship Foundation.

\end{document}